\documentstyle[preprint,psfig,aps]{revtex}
\draft
             
\begin{document}                

\def\be{\begin{equation}}
\def\ee{\end{equation}}
\def\ba{\begin{eqnarray}}
\def\ea{\end{eqnarray}}


\title{Yrast line for weakly interacting trapped bosons}
\author{George F. Bertsch and Thomas Papenbrock}
\address{Institute for Nuclear Theory, Department of Physics, 
University of Washington, Seattle, WA 98195, USA}
\maketitle
\begin{abstract}
We compute numerically the yrast line for harmonically trapped boson systems
with a weak repulsive contact interaction, studying the transition to a vortex
state as the angular momentum $L$ increases and approaches $N$, the number of
bosons.  The $L=N$ eigenstate is indeed dominated by particles with unit
angular momentum, but the state has other significant components beyond the
pure vortex configuration.  There is a smooth crossover between low and high
$L$ with no indication of a quantum phase transition.  Most strikingly, the
energy and wave function appear to be analytical functions of $L$ over the
entire range $2\le L \le N$. We confirm the structure of low-$L$ states
proposed by Mottelson, as mainly single-particle excitations with two or three
units of angular momentum.
\end{abstract}
The low-lying excitations of atomic Bose Einstein condensates in harmonic traps
\cite{Anderson,Bradley,Ketterle} are of considerable experimental and
theoretical interest \cite{Stringari}. Recently, Mottelson proposed a theory 
for the yrast line of weakly interacting $N$-boson systems
\cite{Mottelson}, i.e. the ground states at nonvanishing angular momentum
$L$. Physical arguments led him to assume that the yrast states are excited
upon acting on the ground state $|0\rangle$ of vanishing angular momentum with
a collective operator $Q_\lambda=\sum_{p=1}^Nz_p^\lambda$ that is a sum
of single-particle operators acting on the coordinates $z_p=x_p+iy_p$ of the
$p^{\rm th}$ particle. For angular momenta $L\ll N$ the yrast states are found
to be dominated by quadrupole ($\lambda=2$) and octupole ($\lambda=3$) modes.
Assuming a vortex structure of the yrast states with $L\approx N$ then led to
the prediction of a quantum phase transition in Fock space when passing from
the low angular momentum regime $L\ll N$ to the regime of high angular momenta
$L\approx N$. The reason for this is behavior is the approximate orthogonality
of the collective states $Q_\lambda|0\rangle$ and the single-particle
oscillator states of the vortex line in the regime $N^{1/2}\ll L$. These
results have been obtained for harmonically trapped bosons with a weak
repulsive contact interaction. The case of an attractive interaction has been
studied by Wilkin {\it et al.} \cite{Wilkin}. In this case the total angular
momentum is carried by the center of mass motion, and there are no excitations
corresponding to relative motion. This is not unexpected since internal
excitations would increase the energy of the yrast state.

It is the purpose of this letter to present an independent numerical
computation of the yrast line and to compare with 
Mottelson's results \cite{Mottelson}. In particular we want to focus on the
transition from low to high angular momentum yrast states. The investigation of
this transition is of interest not only for the physics of Bose-Einstein
condensates. Localization in Fock space is under investigation also in
molecular \cite{Wolynes} and condensed matter physics \cite{Altshuler,Carlos}.
The numerical computation has the advantage that it does not rely on the
assumptions made in the analytical calculation. However, with our numerical
methods it is limited to angular momenta below about $\L\approx 50$. 
Most interestingly, our numerical results suggest that the yrast line and the 
corresponding wave functions can be presented by  rather simple analytical
expressions. 

Let us consider $N$ bosons in a two-dimensional harmonic trap interacting via a
contact interaction\footnote{The results obtained below extend to the three
dimensional problem for $L=L_z$.}. We are interested in the
yrast line in the perturbative regime of weak interactions.  Note however that
experimental studies of trapped condensates are often in a regime where the
interaction energy is comparable to the trapping potential, and this may
introduce qualitatively different physics.  We write the Hamiltonian as \be
\label{ham}
\hat{H}=\hat{H}_0 + \hat{V}.
\ee
Here 
\be
\label{H0}
\hat{H}_0=\hbar\omega \sum_j j \,\hat{a}_j^\dagger \hat{a}_j
\ee
is the one-body oscillator Hamiltonian and 
\be
\label{Hint}
\hat{V}=g\sum_{i,j,k,l}V_{ijkl}\,\hat{a}_i^\dagger \hat{a}_j^\dagger \hat{a}_k
\hat{a}_l.  
\ee 
the two-body interaction. The operators $\hat{a}_m$ and $\hat{a}_m^\dagger$
annihilate and create one boson in the single-particle oscillator state
$|m\rangle$ with energy $m\hbar\omega$ and angular momentum $m\hbar$,
respectively and fulfill bosonic commutation rules. The ground state energy is
set to zero. Up to some irrelevant overall constant the matrix elements 
are given by
$V_{ijkl}=2^{-k-l}(k+l)!/(i!j!k!l!)^{1/2}$ and vanish for $i+j\ne
k+l$. For total angular momentum $L$ the Fock space is spanned by states
$|\alpha\rangle\equiv|n_0,n_1,\ldots,n_k\rangle$ with $\sum_{i=0,k}n_i=N$,
$\hat{a}_j^\dagger\hat{a}_j|n_0,n_1,\ldots,n_k\rangle
=n_j|n_0,n_1,\ldots,n_k\rangle$ and $\sum_{j=0,k}j n_j=L$. Here $n_j$ denotes
the occupation of the $j^{\rm th}$ single particle state $|j\rangle$. For
vanishing coupling $g$ the basis states are degenerate in energy, and the
problem thus consists in diagonalizing the two-body interaction $\hat{V}$
inside the Fock space basis. To set up the matrix we act with the operator
(\ref{Hint}) onto one initial basis state with angular momentum $L$ and onto
all states created by this procedure until the the Fock space is exhausted
\cite{PB}. The resulting matrix is sparse, and the yrast state is computed using
a Lanczos algorithm \cite{Arpack}. We restrict ourselves to $L\le 50$
corresponding to a maximal Fock space dimension of about $d_L\approx 2\cdot
10^5$.

\begin{figure}
  \begin{center}
    \leavevmode
    \parbox{0.9\textwidth}
           {\psfig{file=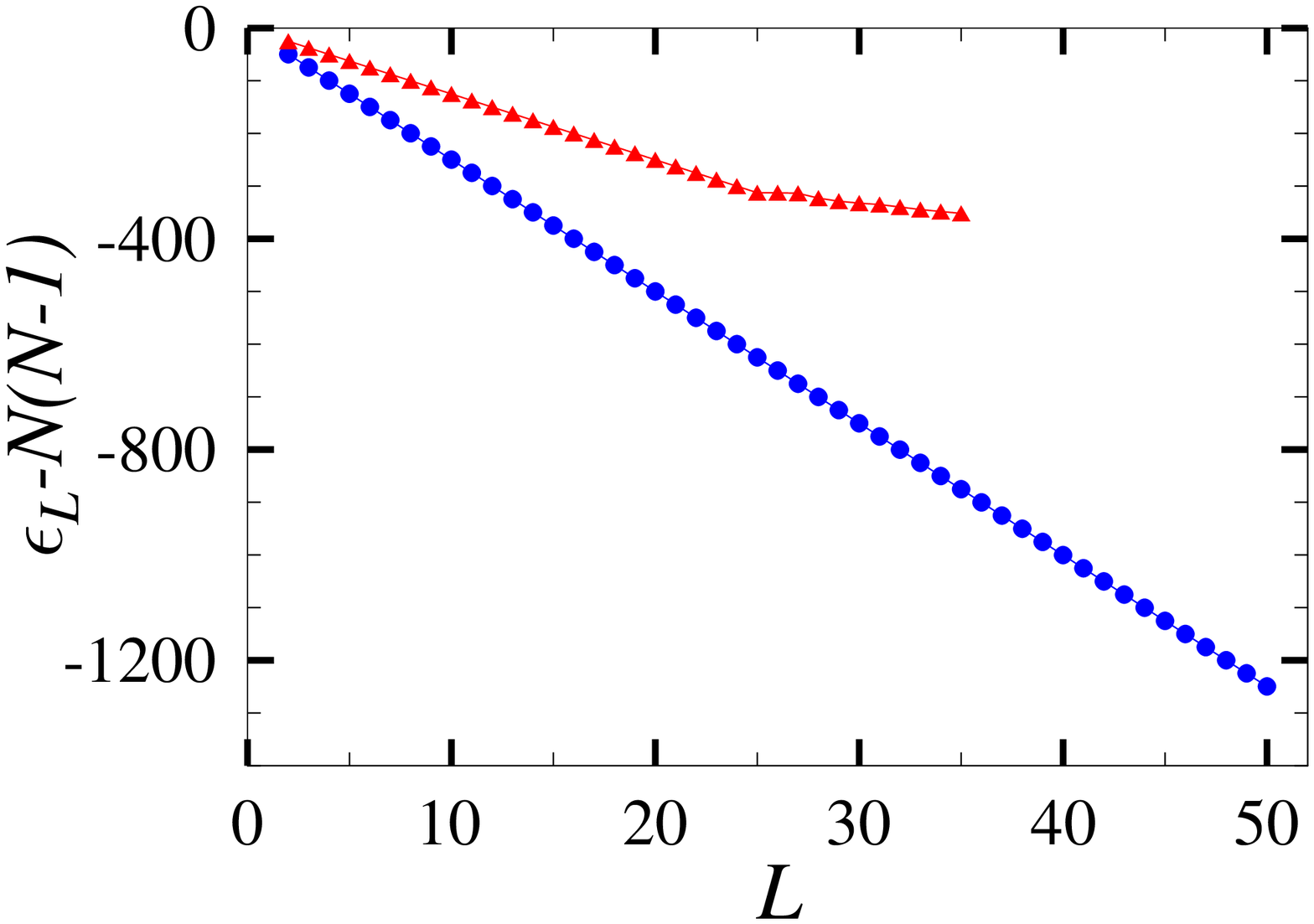,width=0.8\textwidth,angle=0}}
  \end{center}
\protect\caption{Interaction energy $\epsilon_L$ as a function of angular
momentum $L$ for systems of $N=50$ (circles) and $N=25$ bosons (triangles).}
\label{fig1}
\end{figure}

The yrast line, i.e. the ground state energies as a function of the angular
momentum may be written as $E(L)=L\hbar\omega + g\epsilon_L$. Fig.~\ref{fig1}
shows the $L$-dependence of the energies $\epsilon_L$ for systems
of $N=$ 25 and 50 bosons. The energies $\epsilon_L$ simply decrease linearly
with increasing angular momentum for $L\le N$.  In fact to machine precision,
the energy function is found to described by an algebraic expression,
\begin{equation}
\epsilon_L = { N ( 2N-L-2)\over 2}
\end{equation}   
At fixed angular momentum $L$ and for $L\ll N$ the energies $g\epsilon_l$
increase as expected with the square of the number of bosons $N$.  Notice in
the figure that there is a kink in the slope at $N=L$.  This is a hint of
condensation into a vortex state: in macroscopic superfluids, the state for
$L=N$ would have a condensate of unit angular momentum and would be lower in
energy than neighboring yrast states.

\begin{figure}
  \begin{center}
    \leavevmode
    \parbox{0.9\textwidth}
           {\psfig{file=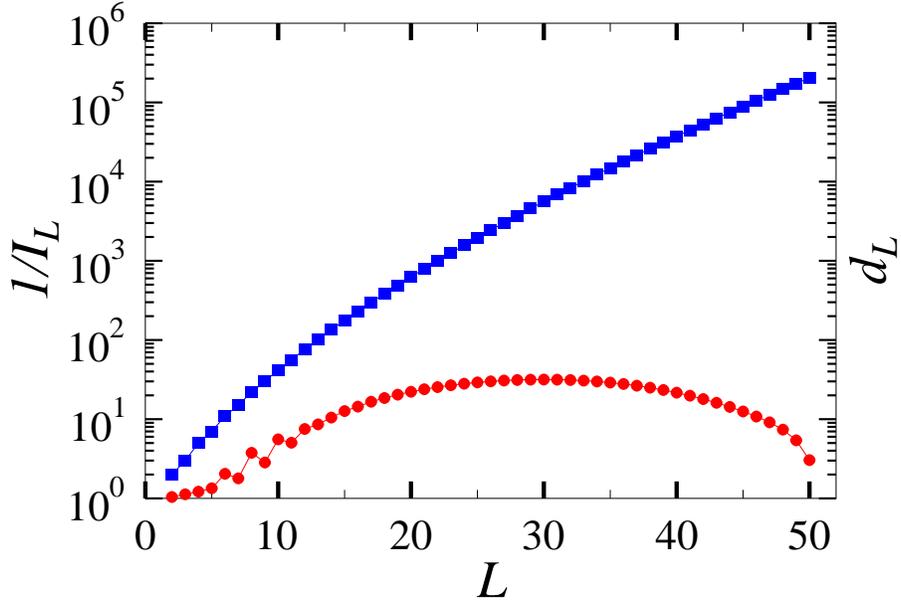,width=0.8\textwidth,angle=0}} 
  \end{center}
\protect\caption{Inverse participation ratio $1/I_L$ (circles) and Fock
    space dimension $d_L$ (squares) as a function of angular momentum $L$ for a
    system of $N=50$ bosons.}
\label{fig2}
\end{figure}

We next investigate the structure of the wave functions of yrast states.  We
would like to know how complex the states are and how well they can be
described by single-particle operators acting on simple states. To address the
question of the complexity of the states in the Fock basis, we take the wave
function amplitudes $c_\alpha^{(L)}$ in the Fock representation of the state
$$|L\rangle=\sum_{\alpha=1}^{d_L} c_\alpha^{(L)} |\alpha\rangle$$
and compute the inverse participation ratio \cite{Kaplan}
$$I_L\equiv \sum_{\alpha=1}^{d_L} |c_\alpha^{(L)}|^4.$$ The $I_L$ is the first
nontrivial moment of the distribution of wave function intensities
$|c_\alpha^{(L)}|^2$. Its inverse $1/I_L$ measures the number of basis states
$|\alpha\rangle$ that have significant overlap with the yrast state
$|L\rangle$.  Fig~\ref{fig2} shows a plot of $1/I_L$ and the Fock space
dimension $d_L$ as a function of angular momentum $L$ for a system of $N=50$
bosons.  The $1/I_L$ is seen to be much smaller than the dimensionality of the
Fock space.  Even where the participating is greatest, at midvalues of $L$,
only about 30 states are active participants. A similar behavior of quantum
non-ergodicity has been found previously in numerical studies \cite{PB}.
Notice that the inverse participation ratio decreases strongly as $N=L$ is
approached. This shows that the yrast state becomes simpler, again hinting at
the formation of a vortex condensate.  Examining the coefficients for $N=25$ in
detail, the largest amplitude at $L=N$ is in fact the vortex state,
$|\alpha\rangle = | 0 N 0 ... 0>$, but it has less than half the probability of
the complete wave function. Interestingly, our numerically obtained yrast state
$|L=N=25\rangle$ agrees with the conjecture given by Wilkin {\it et al.}
\cite{Wilkin}, i.e.  $|L=N\rangle=\prod_{p=1}^N(z_p-z_c)\,|0\rangle$ with
$z_c=N^{-1}\sum_{p=1}^N z_p $ being the center of mass. Based on our numerical
wave functions, we can generalize this conjecture. We believe that all the
yrast states for $2\le L\le N$ are given by the formula
\be
\label{MASTER}
|L\rangle = \sum_{p_1<p_2<\ldots<p_L} (z_{p_1}-z_c)(z_{p_2}-z_c)\ldots
(z_{p_L}-z_c)\,|0>  
\ee 
We have verified that this formula is correct (up to machine precision) by
comparison with the numerically obtained yrast states for $N=25$.  Since the
operator acting on the ground state is translationally invariant no quanta of
the center of mass motion are excited. Notice that there is a natural
termination of the construction at $L=N$.

\begin{figure}
  \begin{center}
    \leavevmode
    \parbox{0.9\textwidth}
           {\psfig{file=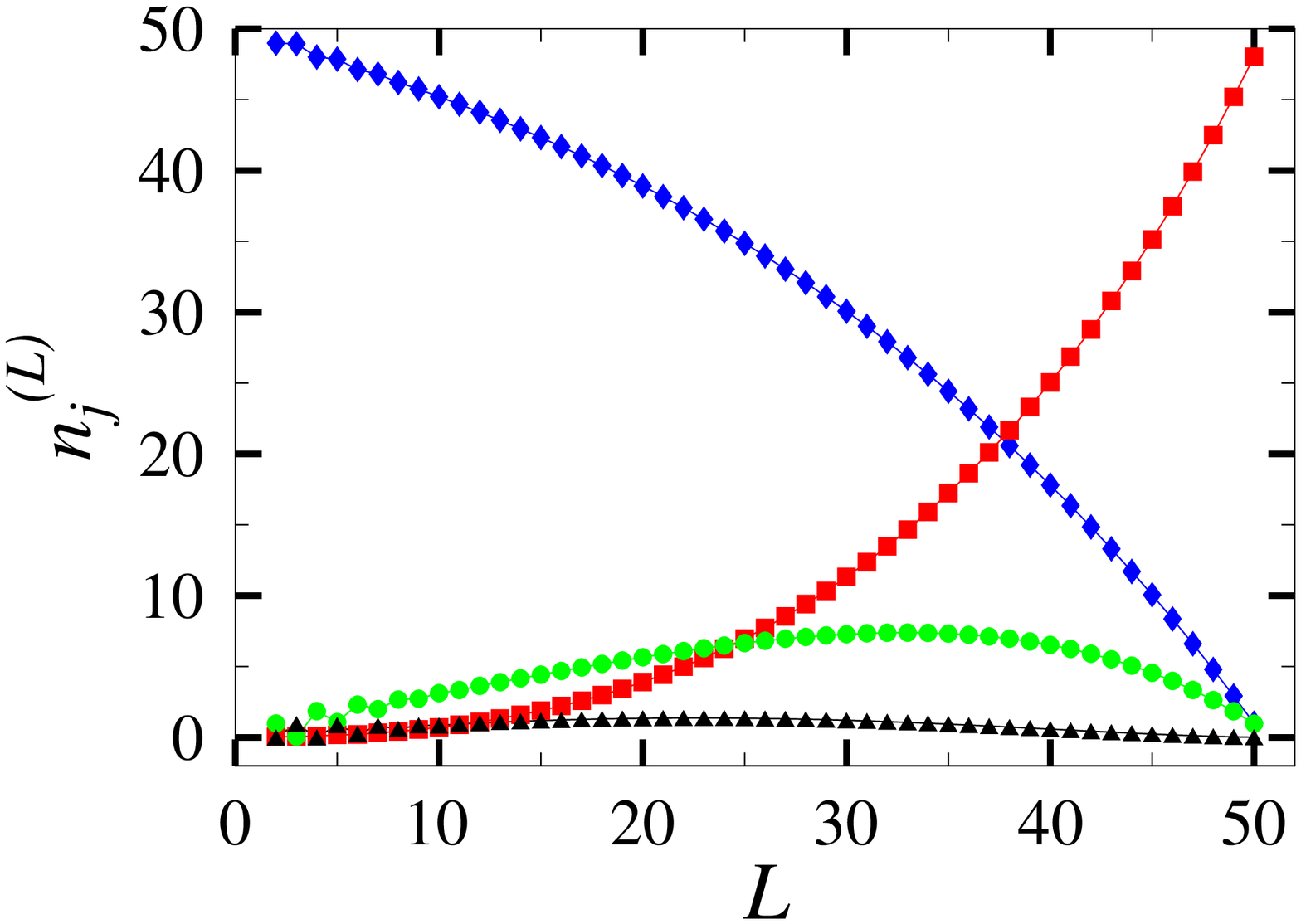,width=0.8\textwidth,angle=0}}
  \end{center}
\protect\caption{Occupation numbers of the lowest single particle
oscillator states as a function of angular momentum $L$ for a system of $N=50$
bosons.($j=0$: diamonds, $j=1$: squares, $j=2$ circles, $j=3$ triangles)}
\label{fig3}
\end{figure}

To further examine the structure of the yrast states $|L\rangle$ we
show a plot of the occupation numbers $n_j^{(L)}\equiv\langle
L|\hat{a}_j^\dagger\hat{a}_j|L\rangle$ for $j=0,1,2,3$ in Fig.~\ref{fig3} for a
system of $N=50$ bosons. At very low angular momenta the yrast states are
dominated by single particle oscillator states with two or three units of
angular momentum. This is in agreement with Mottelson's results
\cite{Mottelson}. However, at larger angular momentum $L$ the dominant fraction
is carried by single particle states with one unit of angular momentum. Note
that the occupation numbers $n_j^{(L)}$ are very small for $j > 3$. This
analysis confirms the results found for the inverse participation ratio.  Note
also that the observables $n_j^{(L)}$ are very smooth functions of $L$. If
there were a quantum phase transition at large $L\approx N/2$, we would expect
to see some precursor in these observables.

In conclusion, our numerical study strongly indicates that there is no quantum
phase transition to a vortex state for trapped condensates in the limit that
the interaction potential is small compared to the oscillator frequency.  The
strongest evidence is the apparent existence of analytic expressions for the
energies and the wave functions on the yrast line for $2\le L\le N$ One might
speculate that these states are contained in a dynamical symmetry group, but we
have no idea how this might come about\footnote{We note also that there is
another symmetry group \cite{pi98}, $SO(2,1)$ , that produces relationships
between energies of different states within a single $L$ subspace.}.  We have
also examined the structure of the yrast states and the matrix elements between
them, finding that the observables vary smoothly with $L$, for $L$ not too
small.

We acknowledge conversations with B. Mottelson.  This work was supported
by the Dept. of Energy under Grant DE-FG-06-90ER40561.

\begin{references}  
%
\bibitem{Anderson}
M. N. Anderson, J. R. Ensher, M. R. Matthews, C. E. Wieman, and E. A. Cornell,
Science {\bf 269}, 198 (1995)
%
\bibitem{Bradley}
C. C. Bradley, C. A. Sacket, J. J. Tollet, and R. G. Hulet,
\prl {\bf 75}, 1687 (1995)
%
\bibitem{Ketterle}
K. B. Davis, M.-O. Mewees, M. R. Andrews, N. J. van Druten, D. S. Durfee, 
D. M. Kurn, and W. Ketterle,
\prl {\bf 75}, 3969 (1995)
%
\bibitem{Stringari}
For a review see, e.g.,
F. Dalfovo, S. Giorgini, L. P. Pitaevskii, and S. Stringari,
\rmp {\bf 71}, 463 (1999)
%
\bibitem{Mottelson}
B. Mottelson,
e-print cond-mat/9905053
%
\bibitem{Wilkin}
N. K. Wilkin, J. M. Gunn, and R. A. Smith,
\prl {\bf 80}, 2265 (1998)
%
\bibitem{Wolynes}
D. M. Leitner and P. G. Wolynes,
Chem. Phys. Lett. {\bf 258}, 18 (1996)
%
\bibitem{Altshuler}
B. L. Altshuler, Y. gefen, A. Kamenev, and L. S. Levitov,
\prl {\bf 78}, 2803 (1997)
%
\bibitem{Carlos}
C. Mej{\'\i}a-Monasterio, J. Richert, T. Rupp, and H. A. Weidenm\"uller,
\prl{\bf 81}, 5189 (1998)
%
\bibitem{PB}
T. Papenbrock and G. F. Bertsch,
\pra {\bf 58}, 4854 (1998)
%
\bibitem{Arpack}
R. B. Lehoucq, D. C. Sorensen, and Y. Yang,
{\it ARPACK User's Guide: Solution to large scale eigenvalue problems with implicitly restarted Arnoldi methods}, 
FORTRAN code available under http://www.caam.rice.edu/software/ARPACK/
%
\bibitem{Kaplan}   
L. Kaplan,
Nonlinearity {\bf 12}, R1 (1999)
%
\bibitem{pi98} L.P. Pitaevshii and A. Rosch, Phys. Rev. A55 R853 (1998).
%
\end{references}
\end{document}